\documentclass[a4paper,11pt]{article}
\pdfoutput=1
\usepackage{jheppub} 
\usepackage{amsmath}
\usepackage{graphicx}
\usepackage{amssymb}
\usepackage{color}    
\usepackage{url}
\usepackage{slashed}
\usepackage{soul}
\usepackage{booktabs,multirow,tabularx}
\usepackage{braket}
\usepackage[table]{xcolor}
\usepackage{tabularx}
\usepackage{array}

\newcommand{\bqa}{\begin{eqnarray}}
\newcommand{\eqa}{\end{eqnarray}}
\newcommand{\beq}{\begin{equation}}
\newcommand{\eeq}{\end{equation}}

\newcommand{\nn}{\nonumber}
\allowdisplaybreaks

\title{Electroweak corrections to Higgs+jet production in gluon fusion}
\author[a]{Long-Bin Chen}
\author[b]{, Hai Tao Li}
\author[c]{, and Wen-Long Sang}

\affiliation[a]{School of Physics and Materials Science, Guangzhou University, Guangzhou 510006, China}

\affiliation[b]{School of Physics, Shandong University, Jinan, Shandong 250100, China}

\affiliation[c]{School of Physical Science and Technology,
Southwest University, Chongqing 400700, China}

\emailAdd{ chenlb@gzhu.edu.cn}
\emailAdd{ haitao.li@sdu.edu.cn~(corresponding author)}
\emailAdd{ wlsang@swu.edu.cn~(corresponding author)}

\preprint{CPTNP-2025-027}

\abstract{
We present the calculation of complete next-to-leading order electroweak corrections to the Higgs boson production in $gg\to g H$ channel. 
We apply the method of differential equations combined with the selection of optimized master integrals to accomplish the calculation of master integrals. We consider three distinct renormalization schemes.
At leading order, the differential distributions and the total cross section show a strong dependence on the renormalization scheme. However, these discrepancies are considerably suppressed once electroweak corrections are taken into account. For $G_\mu$ scheme, the electroweak  correction amounts to approximately $4.3\%$ of the total cross section. Importantly,  we find that the EW corrections exhibit a strong dependence on Higgs transverse momentum. 
}

\begin{document}

\maketitle
\flushbottom

\section{Introduction}

The Higgs mechanism \cite{Higgs:1964pj,Englert:1964et} plays a central role in the Standard Model (SM) of particle physics, providing a consistent explanation for the origin of mass for elementary particles. 
Since the discovery of the Higgs boson by the ATLAS and CMS collaborations~\cite{ATLAS:2012yve,CMS:2012qbp} marked a milestone in particle physics, a major focus of the LHC program has been the precise measurement of its production and decay channels, which agree well with the Standard Model~\cite{ATLAS:2022vkf,CMS:2022dwd}.

Precision predictions for Higgs boson production are essential for testing the SM and probing for potential signs of new physics at the energy and intensity frontiers. In particular, the production of a Higgs boson in association with a high-$p_T$ jet provides a clean and sensitive channel to study the Higgs kinematics and couplings, and plays a crucial role in differential analyses at the LHC. Experimentally, differential distributions such as the Higgs transverse momentum $p_{T}$ have been measured with high precision by ATLAS and CMS \cite{ATLAS:2023jdk,ATLAS:2023hyd,ATLAS:2022yrq,ATLAS:2022qef,ATLAS:2022fnp, ATLAS:2021tbi,ATLAS:2020wny,CMS:2025fwn,CMS:2025ihj, CMS:2024jbe, CMS:2024ddc,CMS:2023gjz, CMS:2022wpo, CMS:2021gxc,CMS:2020zge, CMS:2018gwt}.   The current experimental uncertainty for Higgs $p_T$ distribution measurement is of order 10-15\%, which is dominated by the statistical error. In the future at the High-Luminosity LHC, the statistical error is expected to be reduced to 2.5\%~\cite{Huss:2025nlt}. If one can improve the current systematic error, we will expect percent level measurements for Higgs $p_T$. 

On the theoretical side, the NLO QCD corrections with full top-quark mass dependence have been computed~\cite{Kudashkin:2017skd, Lindert:2018iug, Jones:2018hbb, Chen:2021azt,Bonciani:2022jmb}. The NLO QCD corrections to the inclusive cross section are about 65\%~\cite{Jones:2018hbb}, and NNLO QCD predictions are available in the large top-mass effective theory limit~\cite{Boughezal:2013uia, Chen:2014gva,Boughezal:2015dra,Chen:2016zka}.  Soft-gluon resummation for high  Higgs $p_T$ was performed~\cite{Huang:2014mca,Becher:2014tsa}.  These studies have significantly improved the theoretical control over inclusive and differential Higgs+jet observables. 

Electroweak (EW) corrections, while smaller in magnitude, become non-negligible at the precision frontier.  The EW corrections to Higgs and jet production have been investigated in~\cite{Bonetti:2020hqh,Gao:2023bll,Davies:2023npk,Haisch:2024nzv}. 
In ref. \cite{Bonetti:2020hqh} the corrections through a loop of light quarks have been considered and in ref.~\cite{Davies:2023npk} the top-induced contribution has been computed as an expansion in $1/m_t$.  
The expansion in the limit $m_t \to \infty$ is expected to be valid only in the small centre-of-mass energy region.
In particular, the work~\cite{Gao:2023bll,Haisch:2024nzv} calculated the trilinear self-coupling  $\lambda_{HHH}$ dependent part of the EW corrections and showed that it is promising to set extra constraints on $\lambda_{HHH}$ at the LHC. 
However, EW corrections to  $Hj$ production involve complicated two-loop diagrams with mixed QCD-EW interactions, and currently, results with full EW corrections and top quark mass dependence are not available yet. 

In this work, we present the first complete calculation of EW corrections to the $gg \to g H$ channel at the LHC, including the full top mass dependence.
The two-loop amplitude is organized into form factors, which contain a large number of scalar integrals. All the scalar integrals are reduced into the master integrals (MIs) using integrate-by-part (IBP) identities with public packages, such as {\tt Kira}~\cite{Klappert:2020nbg}, {\tt FIRE} \cite{  Smirnov:2014hma}, and {\tt Blade} \cite{Guan:2024byi}. 
Given that these integrals involve complicated multi-scale elliptic function structures, an analytical computation remains infeasible with existing methodologies, to the best of our knowledge.
We develop a method to select the optimized basis that makes it more efficient to calculate the MIs numerically. After renormalization, we generate the grids for matrix element square at NLO EW corrections. The differential cross sections for Higgs$+$jet production are obtained. 

This paper is organized as follows. The next section presents the form factors and projectors to deal with the amplitudes. The details about the loop calculations and EW renormalization are briefly discussed.  In section~\ref{sec:numc} the inclusive and differential cross sections are shown. Finally, we conclude in section~\ref{sec:concl}.

\section{Amplitudes}

\subsection{General structure~\label{sec:projector}}

\begin{figure}[ht]
\centering
\includegraphics[width=0.9\textwidth]{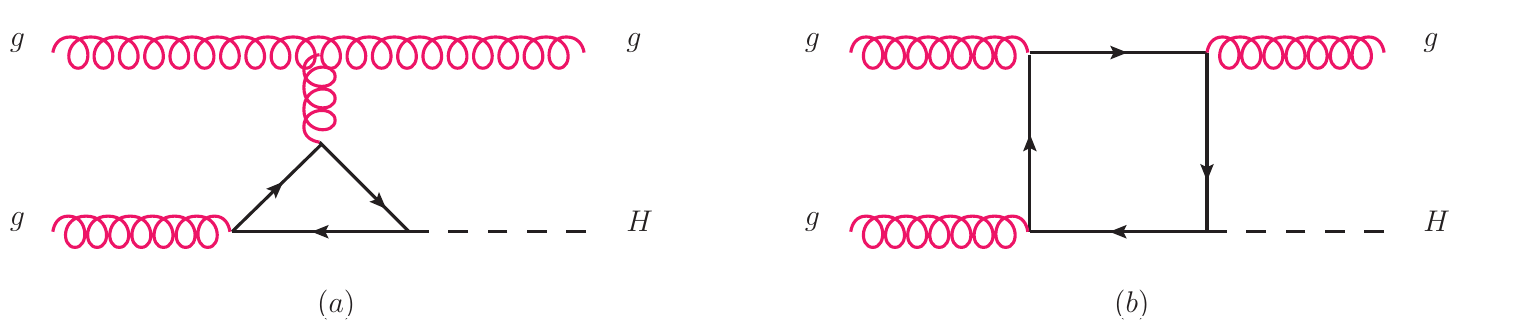}
\caption{ Representative one-loop Feynman diagrams for  $gg \to gH$.}
\label{fig:lod}
\end{figure}

\begin{figure}[ht]
\centering
\includegraphics[width=0.9\textwidth]{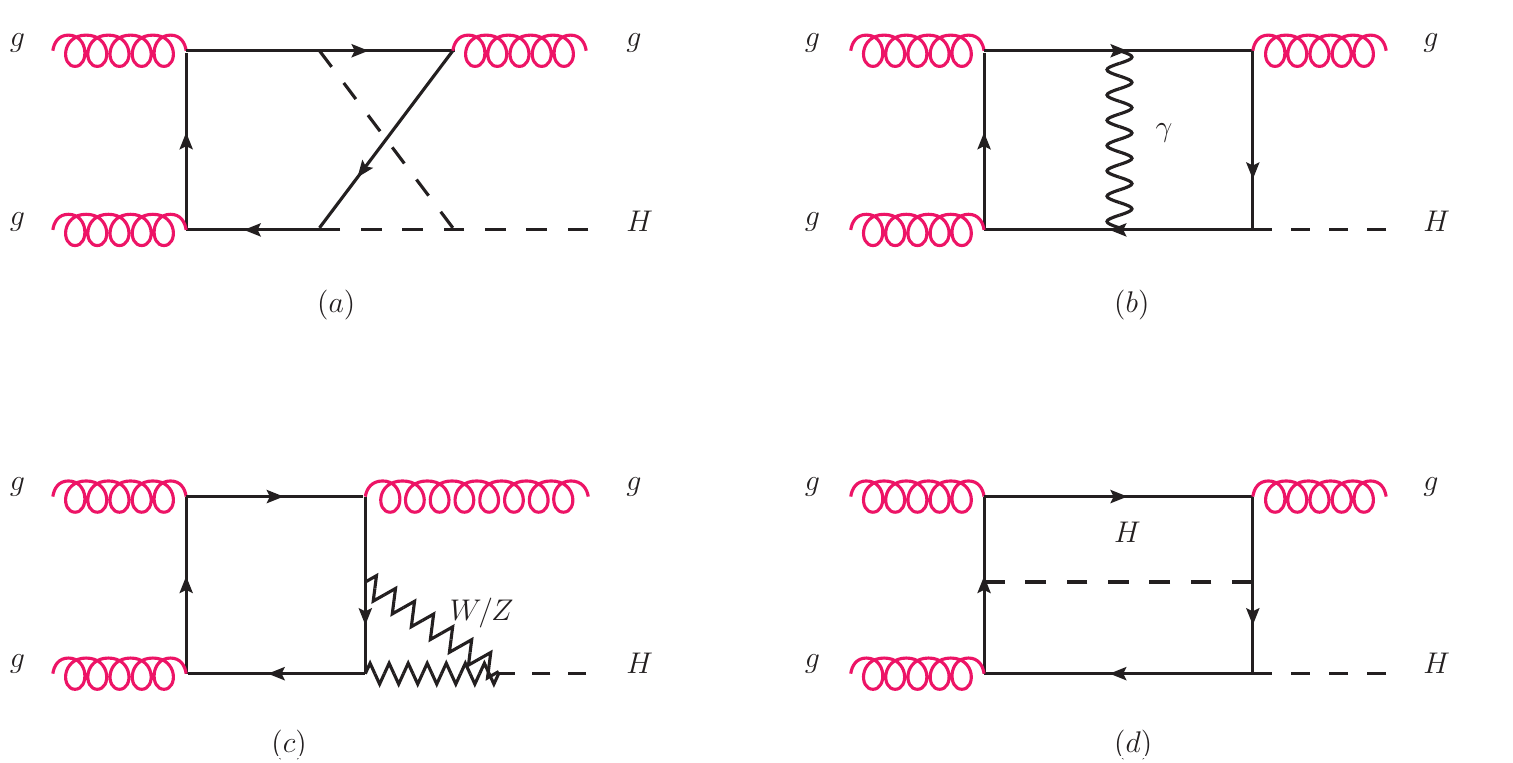}
\caption{Representative  two-loop  Feynman diagrams for EW corrections to  $gg \to g H$.}
\label{fig:nloEW}
\end{figure}

We consider Higgs-boson production in association with a jet at the LHC and focus on the dominant partonic channel, gluon–gluon fusion,
\bqa
g(p_1)\, g(p_2) \to g(p_3) \, H(p_4).
\eqa
 At leading order (LO), the Higgs boson is produced through top-quark loop. The selected LO diagrams are presented in Fig.~\ref{fig:lod}. In this work, we are going to calculate the EW corrections with two-loop Feynman diagrams, which are shown in Fig.~\ref{fig:nloEW}.

The Mandelstam variables for this process are defined as 
\bqa
s=(p_1+p_2)^2,\quad t=(p_1-p_3)^2, \quad u=(p_2-p_3)^2,
\eqa
with the constraint $s+t+u=m_H^2$. The transverse momentum of the Higgs boson can be given by  $p_T^2= t~u/s$.

The one-loop and two-loop amplitudes for $gg\to g H$ process can be written as 
\begin{align}
      M^{abc} = f^{abc} \epsilon_{1,\mu}   \epsilon_{2,\nu}  \epsilon_{3,\rho}^* A^{\mu\nu\rho}(s,t,u),  
\end{align}
and the tensor 
$A^{\mu \nu \rho}(s,t,u)$\footnote{The structure functions also depend on the particle mass, but they are not shown here for brevity.} is a rank-3 tensor under Lorentz transformations. Note that all the Lorentz indices are treated in $d$ dimensions. 
The gauge invariance of the external gluons will impose constraints on the structure functions.
Following the conventions introduced in~\cite{Melnikov:2016qoc}, we choose  the gluons to be transversely polarized and fix the gauge of external gluons as
\begin{align}
\epsilon_j \cdot p_j = 0\,, \quad \epsilon_1 \cdot p_2 = \epsilon_2 \cdot p_3 = \epsilon_3^* \cdot p_1 = 0\,. \label{eq:gauge}
\end{align}
Then, $A^{\mu \nu \rho}(s,t,u)$ can be decomposed into four different Lorentz structures~\cite{Melnikov:2016qoc}
\begin{align}
A^{\mu \nu \rho}(s,t,u) &= F_1(s,t,u) g^{\mu \nu} p_2^\rho 
 + F_2(s,t,u) g^{\mu \rho} p_1^\nu \nonumber \\
&+ F_3(s,t,u) g^{\nu \rho} p_3^\mu 
 + F_4(s,t,u) p_3^\mu p_1^\nu p_2^\rho\,, \label{eq:ff}
\end{align}
with the Lorentz-invariant form factor  $F_j(s,t,u)$.   

With the projector $P_j^{\mu \nu \rho}$, the  form factor  $F_j(s,t,u)$ can be obtained through   
\begin{align}
\sum_{pol}\, P_j^{\mu \nu \rho}\, 
(\epsilon_1^{\mu})^* \epsilon_1^{\mu_1}\;
(\epsilon_2^{\nu})^* \epsilon_2^{\nu_1}\;
\epsilon_3^{\rho} (\epsilon_3^{\rho_1})^*\;A_{\mu_1 \nu_1 \rho_1}(s,t,u)  = F_j(s,t,u)\,.
\label{eq:projdef}
\end{align}
From eq.~\eqref{eq:gauge},  the polarization sums for external gluons are
\begin{align}
&\sum_{pol} \left( \epsilon_1^{\mu}(p_1) \right)^* \epsilon_1^{\nu}(p_1) = 
- g^{\mu \nu} + \frac{p_1^\mu p_2^\nu + p_1^\nu p_2^\mu}{p_1 \cdot p_2}\,, \label{eq:polsums1}\\
&\sum_{pol} \left( \epsilon_2^{\mu}(p_2) \right)^* \epsilon_2^{\nu}(p_2) = 
- g^{\mu \nu} + \frac{p_2^\mu p_3^\nu + p_2^\nu p_3^\mu}{p_2 \cdot p_3}\,,  \label{eq:polsums2}\\
&\sum_{pol}  \epsilon_3^{\mu}(p_3)  \left(\epsilon_3^{\nu}(p_3)\right)^* = 
- g^{\mu \nu} + \frac{p_1^\mu p_3^\nu + p_1^\nu p_3^\mu}{p_1 \cdot p_3} \,. \label{eq:polsums3}
\end{align} 
From the gauge invariance of external gluons and Lorentz invariance, the projector can be written as a linear combination of the following structures
\begin{align}
P_j^{\mu \nu \rho} &= \frac{1}{d-3} 
\left [ c_1^{(j)}\, g^{\mu \nu}\, p_2^\rho
 + c_2^{(j)}\, g^{\mu \rho}\, p_1^\nu + c_3^{(j)}\, g^{\nu \rho}\, p_3^\mu
 + c_4^{(j)}\, p_3^\mu p_1^\nu p_2^\rho\, \right], 
\end{align}
with $c_i^{(j)}$ the function of kinematic invariants and    $ j \in \{1,2,3,4 \}$.  And the coefficients $c_i^{(j)}$ can be obtained by solving linear equations indicated in eq.~\eqref{eq:projdef}, which are   
\begin{equation}\begin{aligned}
c_1^{(1)} &= \frac{t}{s\,u}\,,&
c_2^{(1)} &= 0\,,&
c_3^{(1)} &= 0\,,&
c_4^{(1)} &= \frac{1}{s\,u}\,, 
\\
c_1^{(2)} &= 0\,,&
c_2^{(2)} &= \frac{u}{s\,t}\,,&
c_3^{(2)} &= 0\,,&
c_4^{(2)} &= \frac{1}{s\,t}\,,
\\
c_1^{(3)} &= 0\,,&
c_2^{(3)} &= 0\,,&
c_3^{(3)} &= \frac{s}{t\,u}\,,&
c_4^{(3)} &= -\frac{1}{t\,u}\,,
\\
c_1^{(4)} &= \frac{1}{s\,u}\,,\quad&
c_2^{(4)} &= \frac{1}{s\,t}\,,\quad&
c_3^{(4)} &= -\frac{1}{t\,u}\,,\quad&
c_4^{(4)} &= \frac{d}{s\,t\,u}\,.
\end{aligned}\end{equation}

\subsection{Computational Framework}

We use {\tt FeynArts}\cite{Hahn:2000kx} to generate the Feynman diagrams and amplitudes for the process under study. In total, there are 12 one-loop and 1,416 two-loop diagrams, with representative two-loop examples shown in Fig.~\ref{fig:nloEW}.  The algebraic manipulation of amplitudes, including Lorentz contractions and Dirac algebra, is performed using {\tt FeynCalc}~\cite{ Mertig:1990an, Shtabovenko:2020gxv, Shtabovenko:2023idz},  {\tt FormLink}\cite{Feng:2012tk}, and {\tt Form 4.0} ~\cite{Vermaseren:2000nd,Kuipers:2012rf}.

Following the projector method described in section \ref{sec:projector}, we extract four independent form factors from the amplitudes. The resulting Feynman integrals are then reduced to linear combinations of MIs using IBP identities, implemented through the packages {\tt Kira}\cite{Klappert:2020nbg}, {\tt Blade}\cite{Guan:2024byi}, and {\tt FIRE}\cite{Smirnov:2014hma}.

For the two-loop contributions, the IBP reduction yields 108 distinct integral families and over 3,600 MIs. A substantial portion of these are multi-scale integrals whose analytic evaluation remains extremely challenging,  and, in most cases, is not yet possible to obtain with current methods.

Consequently, we resort to a numerical treatment for the encountered MIs for various phase space points. A standard approach for computing multi-loop, multi-scale Feynman integrals is to construct differential equations~\cite{Kotikov:1990kg} for the MIs with respect to external momenta and masses, and then solve them either analytically or numerically. This method was later generalized to multi-leg processes~\cite{Gehrmann:1999as}, and has become a cornerstone of modern multi-loop calculations. However, for the class of processes considered in this work, the complexity of the kinematics renders this approach highly cumbersome in practice. In particular, transforming the system into a canonical form~\cite{Henn:2013pwa} proves to be extremely challenging or even infeasible in this case.

As an alternative, we construct an optimized basis of MIs that markedly improves the numerical stability and efficiency of solving the differential equations.  This basis is chosen so that every coefficient arising from IBP reduction contains only good denominators, i.e., expressions that are decomposable into polynomials of kinematic invariants and masses, independent of the space-time dimension $d$, and at most linear in $d$ (e.g., terms of the form 
$a\, d+b$, with rational $a$ and $b$).

To identify the optimized basis, one must perform tentative reductions and obtain a set of MIs generated by reduction packages such as {\tt Kira} \cite{Klappert:2020nbg}. However, tentative reductions could be rather slow when bad denominators emerge. To accelerate the search for an optimized basis, we first set kinematic variables (e.g., $s, t, m_W, m_t, m_H$) to large prime values and perform the tentative reduction. When bad denominators appear, the factorized denominators of the reduction coefficients may contain terms with very large coefficients for \(d\). This allows us to identify all sectors containing bad denominators more efficiently. Then within the algorithm similar to \cite{Smirnov:2020quc,Usovitsch:2020jrk}, we successfully identified optimized bases for all 108 integral families. Importantly, potential issues such as spurious or bad denominators are absent in the final physical results, ensuring both consistency and numerical stability.   

Once the MIs are selected, we construct their differential equations with respect to the kinematic variables $s$ and $t$. By contrast, we find that the size of the differential equations for certain complex families, when using an optimized basis, can be an order of magnitude simpler than when using the preselected basis from reduction programs. This will significantly improve the efficiency of numerically solving differential equations.  The boundary conditions for all MIs are evaluated numerically using {\tt AMFlow} \cite{Liu:2017jxz,Liu:2022chg}. We then solve the differential equations numerically to obtain the MIs at arbitrary phase-space points. Employing the optimized basis accelerates solving the differential equations numerically and markedly enhances stability.  We have validated this procedure by comparing its predictions against direct evaluations obtained with {\tt AMFlow} at numerous representative phase-space points; agreement is achieved to high numerical precision, confirming both the robustness and accuracy of our method.

\subsection{EW Renormalization}

We employ the on-shell renormalization scheme~\cite{Ross:1973fp,Hollik:1988ii}, a widely adopted approach in the literature, where the renormalized parameters are chosen to match precisely measured quantities such as the Higgs boson mass, $W/Z$ boson masses, top quark mass, and QED charge coupling.
Specifically, we utilize the {\it Fleischer-Jegerlehner tadpole scheme}~\cite{Fleischer:1980ub,Dittmaier:2022maf,Denner:2019vbn}. That is, instead of introducing a tadpole counterterm, we include all tadpole diagrams in both the bare Feynman diagrams and other Feynman diagrams needed for the counterterms.  

In addition, we adopt three different renormalization schemes for the QED charge coupling: $\alpha(0)$ scheme, $\alpha(M_Z)$ scheme, and $G_\mu$ scheme.  
In the $\alpha(0)$ scheme, the Thomson-limit value is used. At one-loop order, $\delta Z_e$ is expressed as
\beq\label{eq:ze:orig}
\delta Z_e|_{\alpha(0)}= {1\over 2} \Pi^{AA}(0)-{s_W\over c_W}
{\Sigma^{A Z}_T(0) \over M_Z^2},
\eeq
where ${\Pi(s)}\equiv \Sigma_T^{AA}(s)/s$ denotes the photon vacuum polarization. Because ${\Pi(s)}$ at small momentum receives non-perturbative hadronic contributions, we may rewrite $\delta{Z_e}$ as
\bqa
\delta{Z_e}\big|_{\alpha(0)} &=& {1\over 2} \Delta{\alpha_{\rm had}^{(5)}}(M_Z)
+{1\over 2}{\rm Re}\, \Pi^{{AA}(5)}(M_Z^2)
\nn\\
&+&  {1\over 2}\Pi^{AA}_{\rm rem}(0)-{s_W\over c_W}
{\Sigma^{A Z}_T(0)\over M_Z^2},
\label{dZe:alpha(0):scheme}
\eqa
where $\Pi^{{AA}(5)}(M_Z^2)$ is the photon vacuum polarization from five massless quarks at momentum transfer
$M_Z^2$.  $\Delta{\alpha_{\rm had}^{(5)}}(M_Z)$ absorbs the low-energy hadronic piece extracted from $R$ values in
low-energy $e^+e^-$ experiments, while
$\Pi^{AA}_{\rm rem}(0)$ collects perturbatively calculable contributions from the $W$ boson, charged leptons and top quark.

The renormalization constant for the other two schemes can be related to $\delta{Z_e}\big|_{\alpha(0)}$ through the relations:
\bqa
\delta{Z_e}\big|_{\alpha(M_Z)}=\delta{Z_e}\big|_{\alpha(0)}- {1\over 2} \Delta\alpha(M_Z),\quad \quad
\delta Z_e|_{G_\mu}=\delta Z_e|_{\alpha(0)}-{1\over 2}\Delta r,
\label{Delta:alpha:split:into:two:terms}
\eqa
where $\Delta\alpha(M_Z)$ and $\Delta r$ are explicitly presented in refs.~\cite{Denner:1991kt}. The corresponding running couplings are
\begin{subequations}
\bqa
\label{dZe:alpha(MZ):scheme}
&& \alpha\left(M_Z\right)=\frac{\alpha(0)}{1-\Delta \alpha\left(M_Z\right)},
\\
\label{dZe:Gmu:scheme}
&& \alpha_{G_\mu}=\frac{\sqrt{2}}{\pi}G_{\mu} M_W^2\left(1-\frac{M_W^2}{M_Z^2}\right).
\eqa
\end{subequations}
Unlike the $\alpha(0)$ scheme, the $\alpha(M_Z)$ and $G_\mu$ schemes resum large (non-)logarithmic corrections from light fermion and top quark loops, improving perturbative convergence.
For more details, we refer the authors to the refs.~\cite{Denner:1991kt,Sun:2016bel,Sang:2024vqk}.

After this renormalization programme, the form factors $F_1\ldots F_4$  emerge ultraviolet-finite at each phase space point and are ready for phenomenology.

\section{Numerical Results}~\label{sec:numc}

The input parameters used for the LO amplitude squared are
\begin{align}
    m_W=80.399~\text{GeV}, &\quad m_Z=91.1876~\text{GeV},\quad  G_{\mu}=1.16581\times 10^{-5}~\text{GeV}^{-2} \,  \\   m_H=&125~\text{GeV},\quad m_t=172.5~\text{GeV}. \nonumber
\end{align}
The two-loop matrix elements for the EW corrections are evaluated numerically. To facilitate efficient phenomenological applications, we construct interpolation grids for the ratio of the squared matrix elements between the NLO EW and LO contributions for the three renormalization schemes mentioned above.
To enhance numerical efficiency and stability during the computation of these ratios, we adopt a slightly simplified mass input, 
\begin{align}
    m_W=80~\text{GeV}, &\quad m_Z=91~\text{GeV}, \quad
    m_t=172~\text{GeV}.  
\end{align}
with all other particle masses and widths set to zero. This approximation may introduce a deviation of subpercent in the determination of the NLO EW corrections, which is negligible for the purposes of our analysis. Note that if the masses are modified, the amplitude and master integrals must be regenerated, which requires about $10^5$ CPU hours to reconstruct the numerical grid. The fine structure constant $\alpha$ in the other two schemes can be calculated from eq.~(\ref{dZe:alpha(MZ):scheme}) and eq.~(\ref{dZe:Gmu:scheme}).
The Cabibbo-Kobayashi-Maskawa mixing matrix is set to diagonal. 
To guarantee the reliability of our results, we evaluated the master integrals with a numerical precision of more than 45 digits.

Before to present numerical results, we perform a comparison between the full and  the  $m_t$ expanded form factors with ref. [37]. We find that the bare and counterterm contributions display good convergence individually. However, after combining them, large numerical cancellations occur, leading to a relatively slower convergence of the final result compared to the separate contributions. For instance, the difference between the full and $m_t$ expanded results is around $10\%$ at $\sqrt{s}=150$ GeV.

We employ CT18NLO \cite{Hou:2019efy} as our default PDF set for both LO and NLO calculations. The running of strong coupling $\alpha_s$ with two-loop accuracy is provided by the {\tt LHAPDF6} library \cite{Buckley:2014ana}. Our default renormalization and factorization scales are chosen as $\mu=\sqrt{p_T^2+m_H^2}$, where $p_T$ is the transverse momentum of the Higgs boson. The events are selected with higgs $p_T$ larger than 20~GeV. 

\begin{table}[h]
\centering
\begin{tabular}{p{1.5cm}<{\centering}  p{1.5cm}<{\centering} p{2.1cm}<{\centering} p{2.5cm}<{\centering}}
\hline
scheme & $\sigma^{\rm LO}$[pb]~&~$\sigma^{\rm NLO-EW}$[pb]& $K$-factor \\
\hline
$G_\mu$ &  7.91 & 8.25  & 1.043 \\
$\alpha(0)$ & 7.56 & 8.11 & 1.073  \\
$\alpha(m_Z^2)$ &  8.04& 8.16 &  1.015\\
\hline
\end{tabular}
\caption{ Total cross section for $pp\to H+jet$ via the dominant $gg\to g H$ channel at $\sqrt{s}=13$ TeV, with and without EW corrections.}
\label{numeric-values}
\end{table}

The total cross sections with three different renormalization schemes are presented in Table~\ref{numeric-values}. 
All schemes yield positive EW corrections: $4.3\%$ in the  
$G_\mu$ scheme, $7.3\%$ in the $\alpha(0)$ scheme and $1.5\%$ in the $\alpha(m_Z)$ scheme.
The difference among these $\alpha$ schemes provides a conventional estimate of the theoretical uncertainty.  
Comparing the LO and NLO results, we see that this scheme–dependent uncertainty is markedly reduced once the full EW corrections are included.

\begin{figure}[ht]
    \centering
    \includegraphics[width=0.32\linewidth]{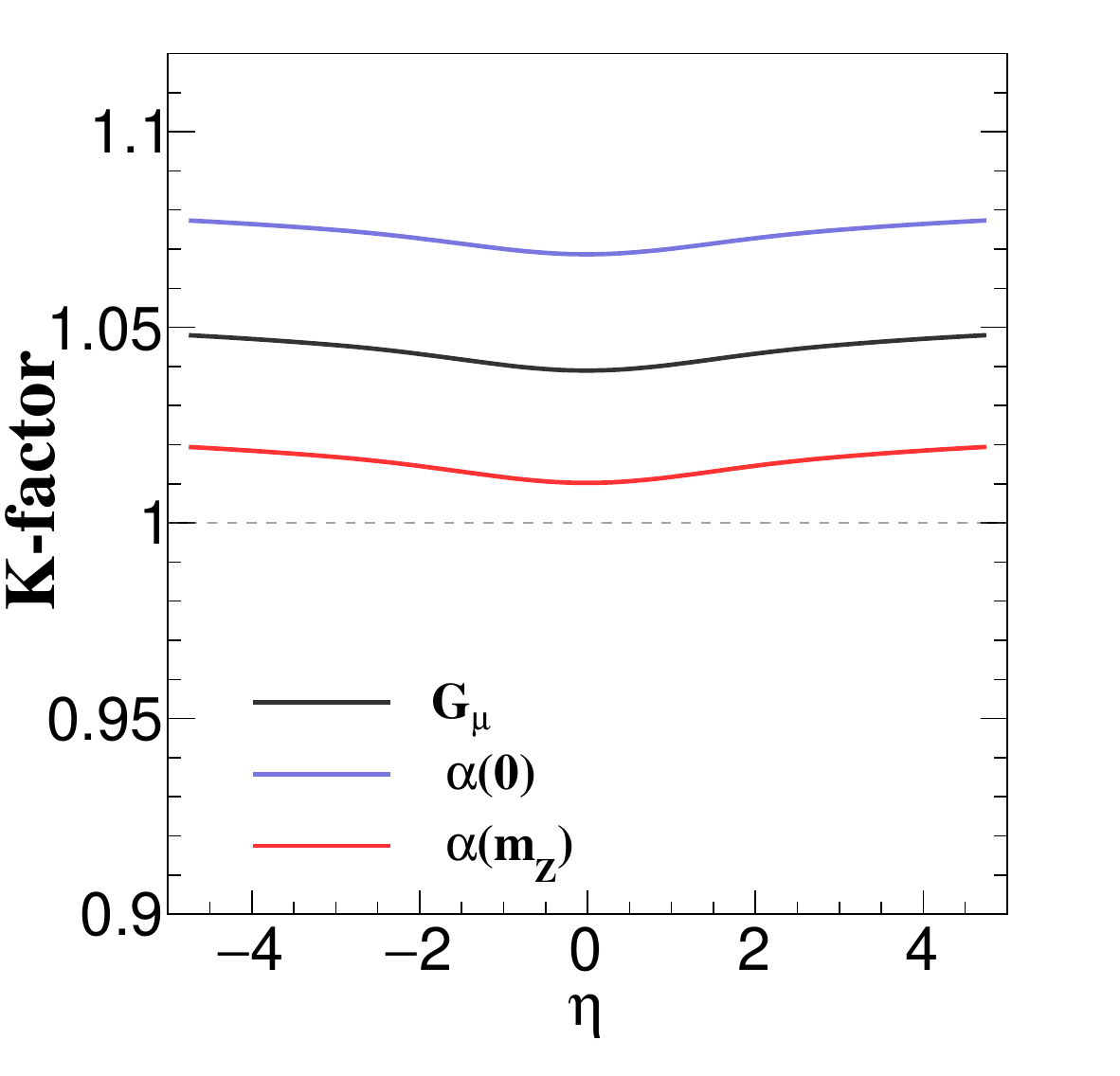}
    \includegraphics[width=0.32\linewidth]{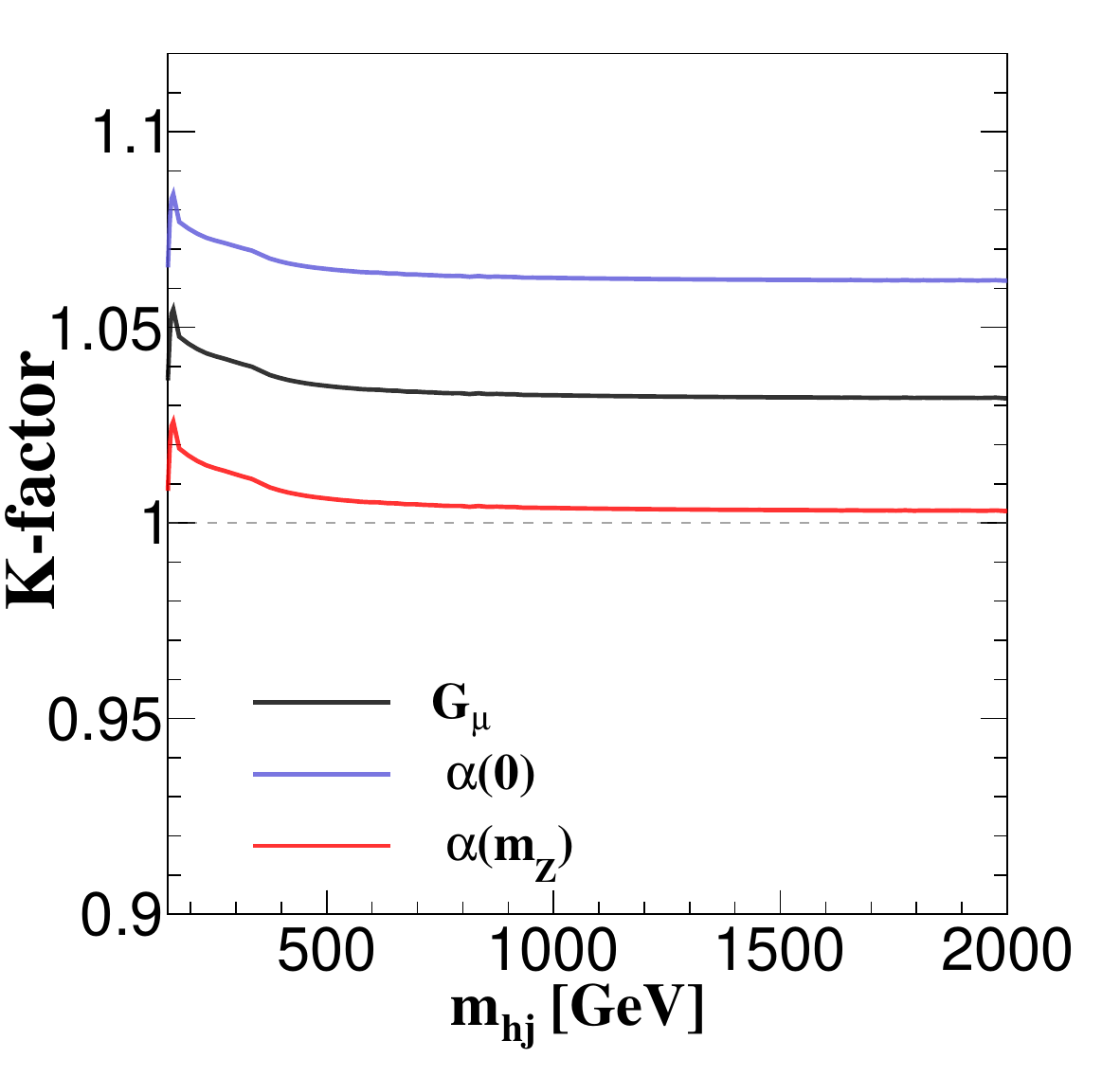}
    \includegraphics[width=0.32\linewidth]{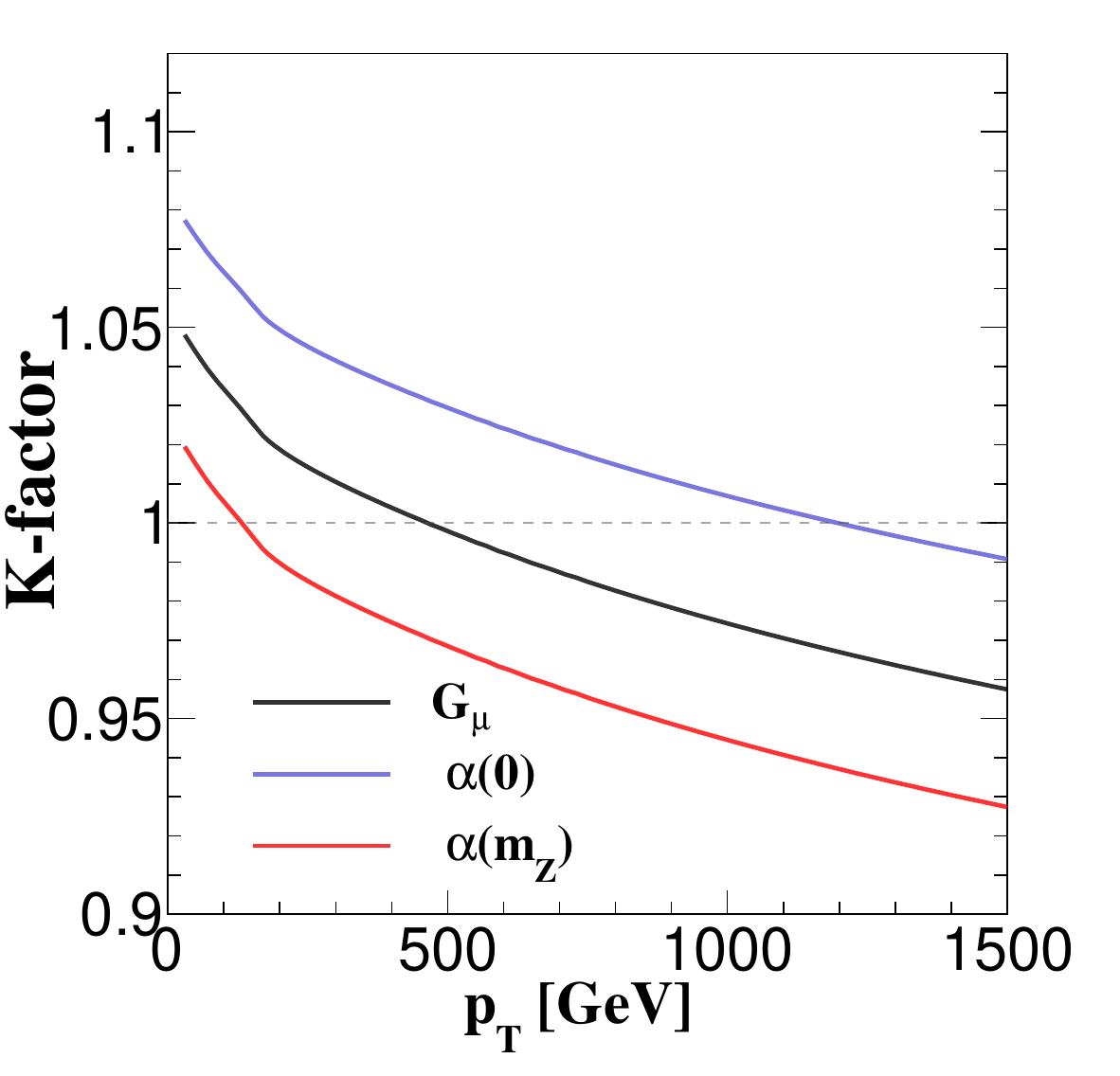}
    \vspace{-0.5cm}
    \caption{Relative EW corrections in $G_\mu$, $\alpha(0)$, and $\alpha(m_Z)$ schemes. The corrections as a function of Higgs pseudorapidity $\eta$, Higgs $p_T$, and $m_{hj}$.  The $K$-factor is $d\sigma_{\rm NLO_{EW}} /d\sigma_{\rm LO} $.  }
    \label{fig:kfactor}
\end{figure}

\begin{figure}[ht]
    \centering
    \includegraphics[width=0.32\linewidth]{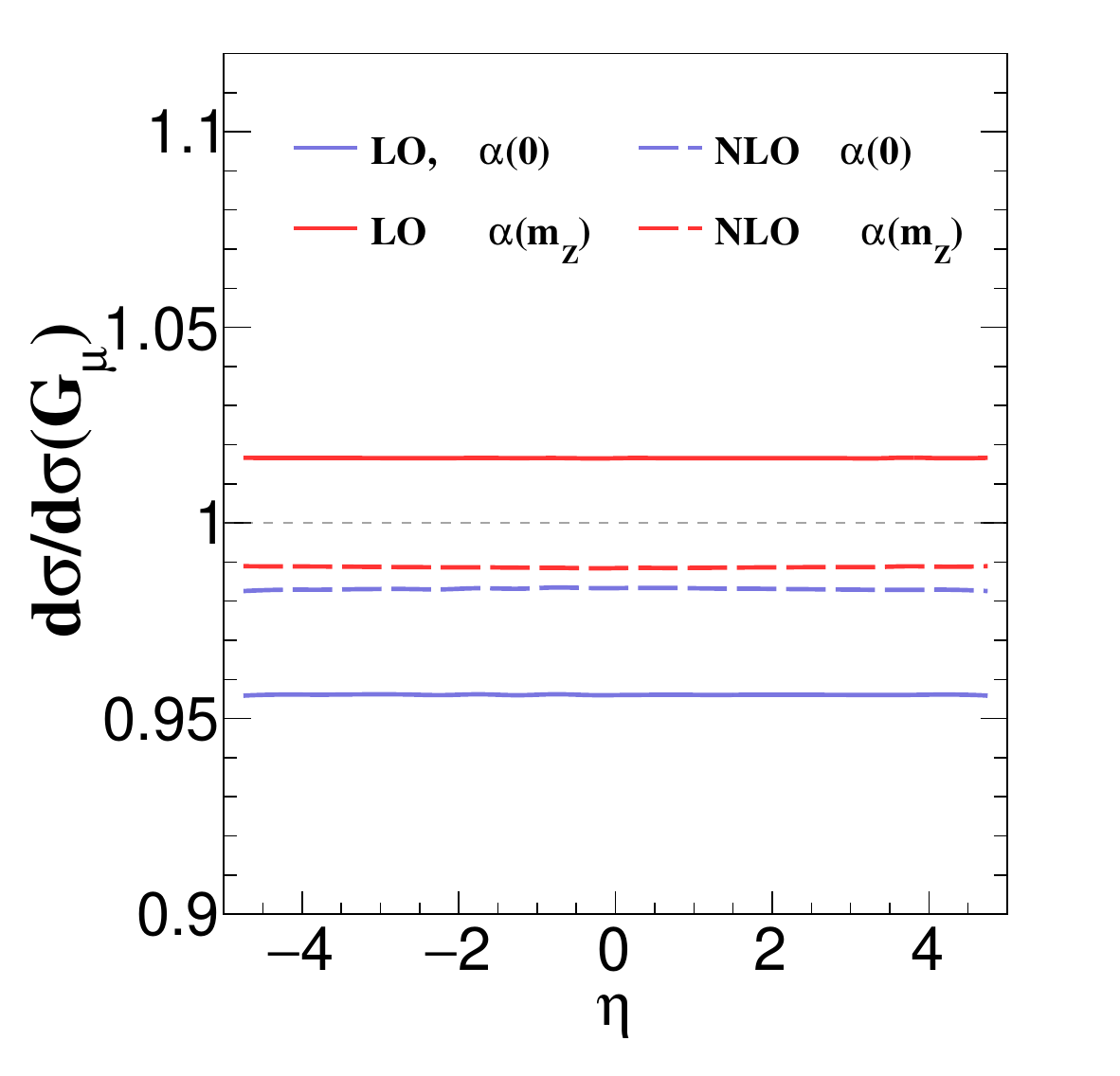}
    \includegraphics[width=0.32\linewidth]{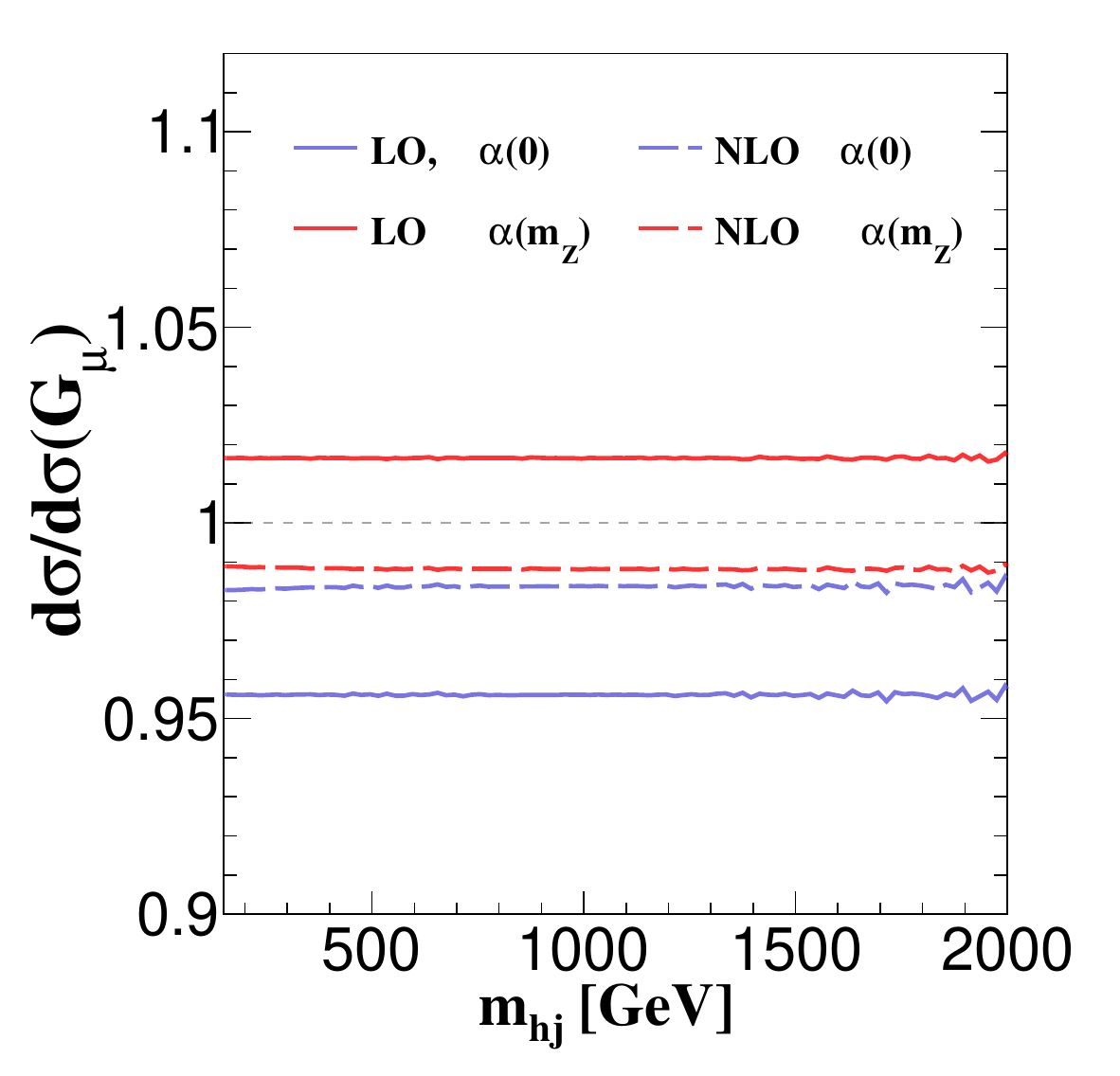}
    \includegraphics[width=0.32\linewidth]{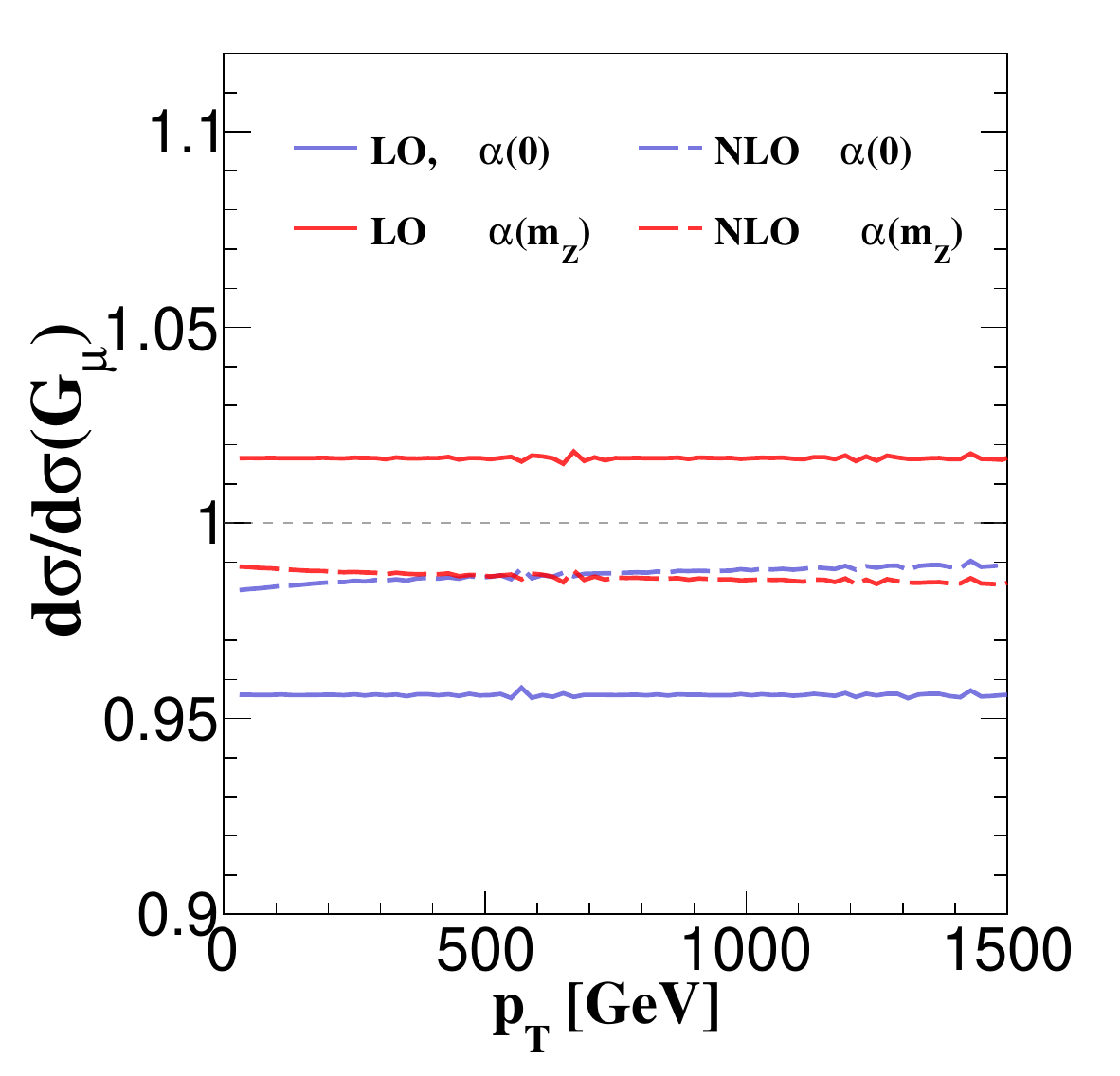}
    \vspace{-0.5cm}
    \caption{Ratios between the distributions in $\alpha(0)$ or $\alpha(m_Z)$ schemes and the ones in $G_{\mu}$ scheme.   }
    \label{fig:ratios}
\end{figure}

In Fig.~\ref{fig:kfactor} we display the differential $K$-factors for the three $\alpha$ schemes. Although each yields corrections of only a few percent, pronounced differences appear, especially in the high-energy regions of the  $p_T$ distributions. 
At lower $p_T$, the absolute size of the EW corrections is largest in the $\alpha(0)$ scheme, smallest in the $\alpha(m_Z)$ scheme, and intermediate in the $G_\mu$ scheme.  However, this ordering is reversed as $p_T$ increases. The $G_\mu$ scheme, absorbing universal corrections to muon decay, shows the relatively stable behaviour across the entire kinematic range.

Figure~\ref{fig:ratios} presents the ratios of the $\alpha(0)$ and $\alpha(m_Z)$ predictions to the $G_{\mu}$ baseline. The LO spread among the schemes is substantial, but it shrinks dramatically once the EW corrections are included, thereby bringing the theoretical uncertainty under much tighter control.

Even though all EW input schemes converge to similar results at NLO, the $G_\mu$ scheme, for which the LO amplitude is explicitly proportional to   $G_\mu$,  remains the most physically motivated choice for this process. Consequently, we adopt the  $G_\mu$ scheme throughout the remainder of our phenomenological study.

\begin{figure}
    \centering
    \includegraphics[width=0.32\linewidth]{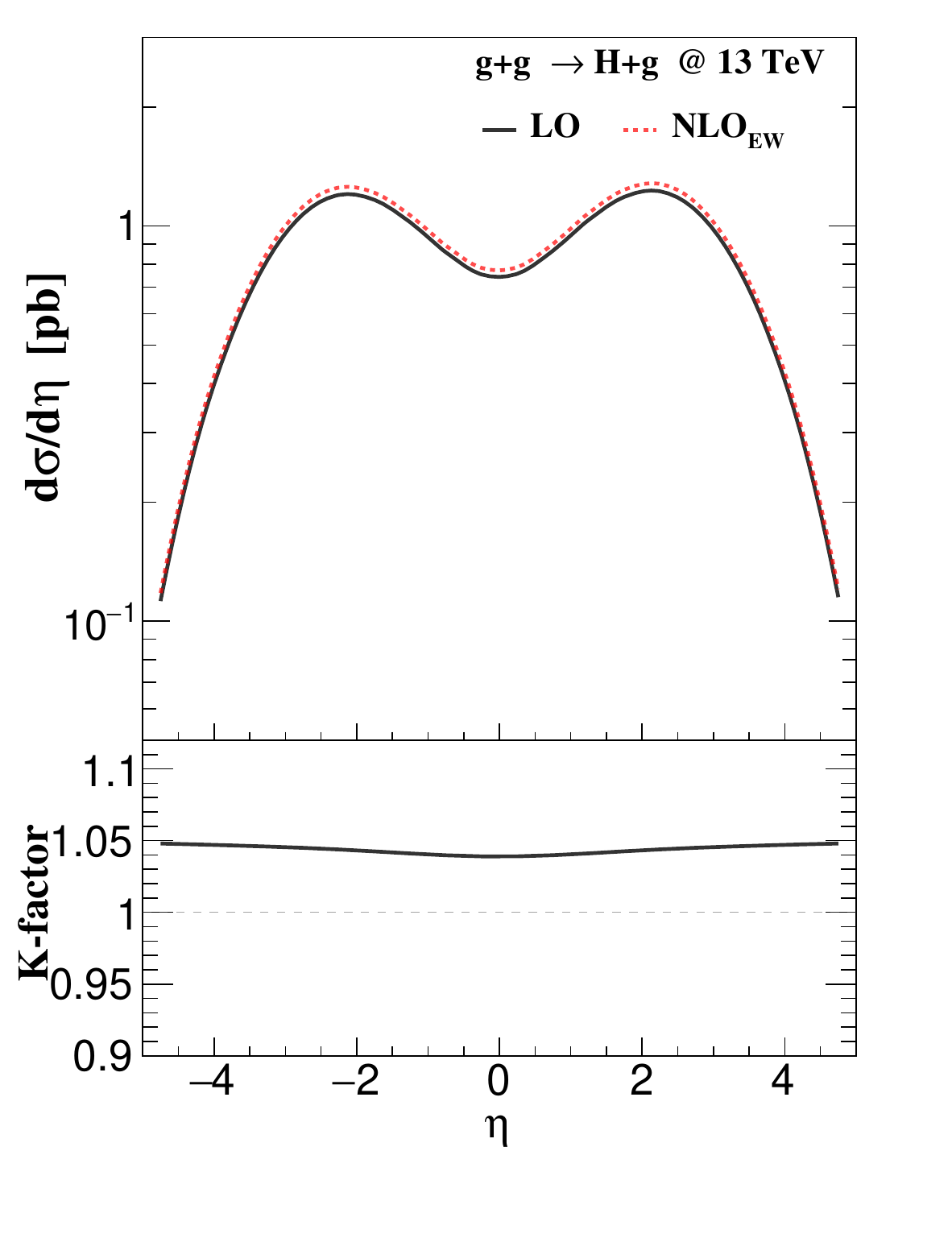}
    \includegraphics[width=0.32\linewidth]{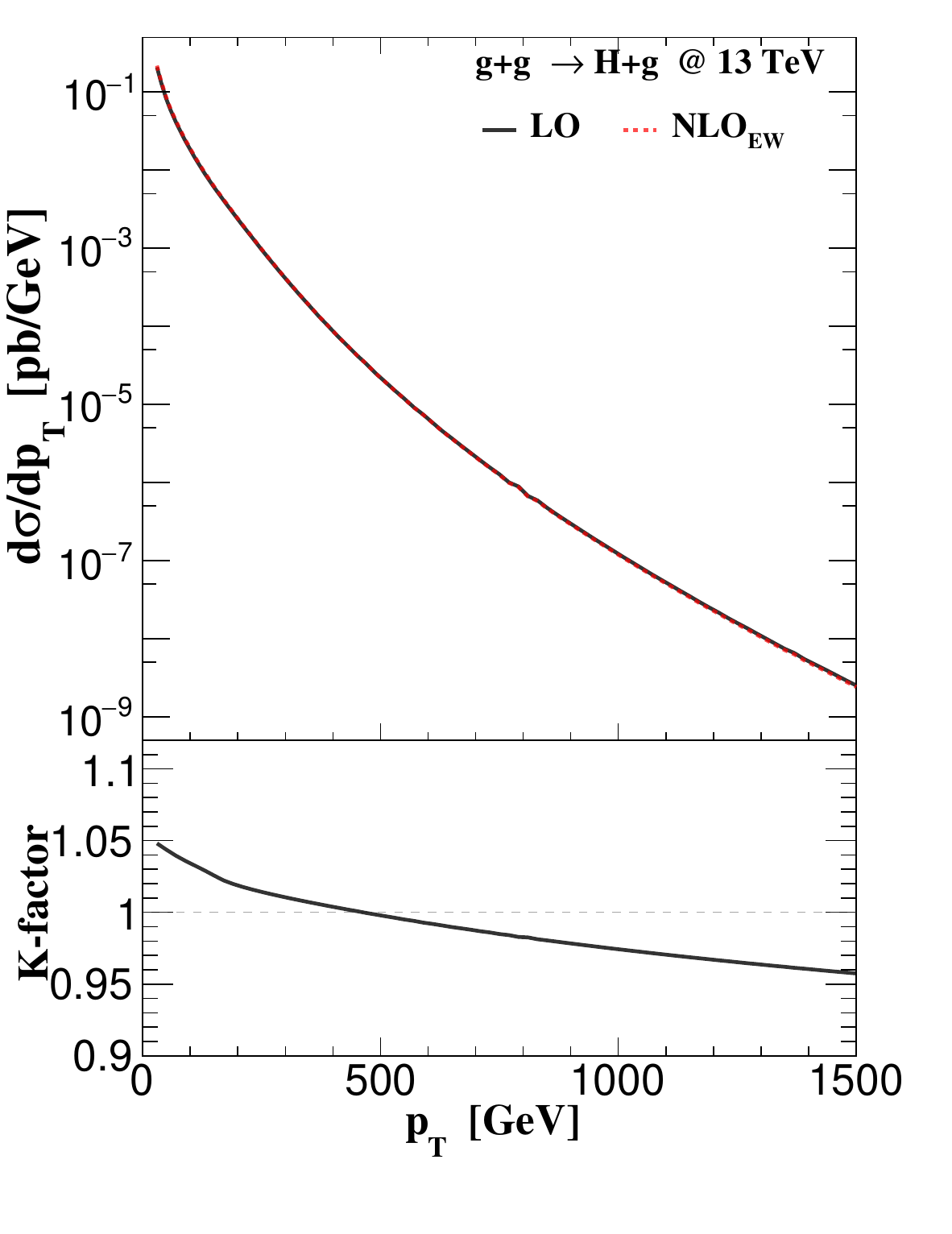}
    \includegraphics[width=0.32\linewidth]{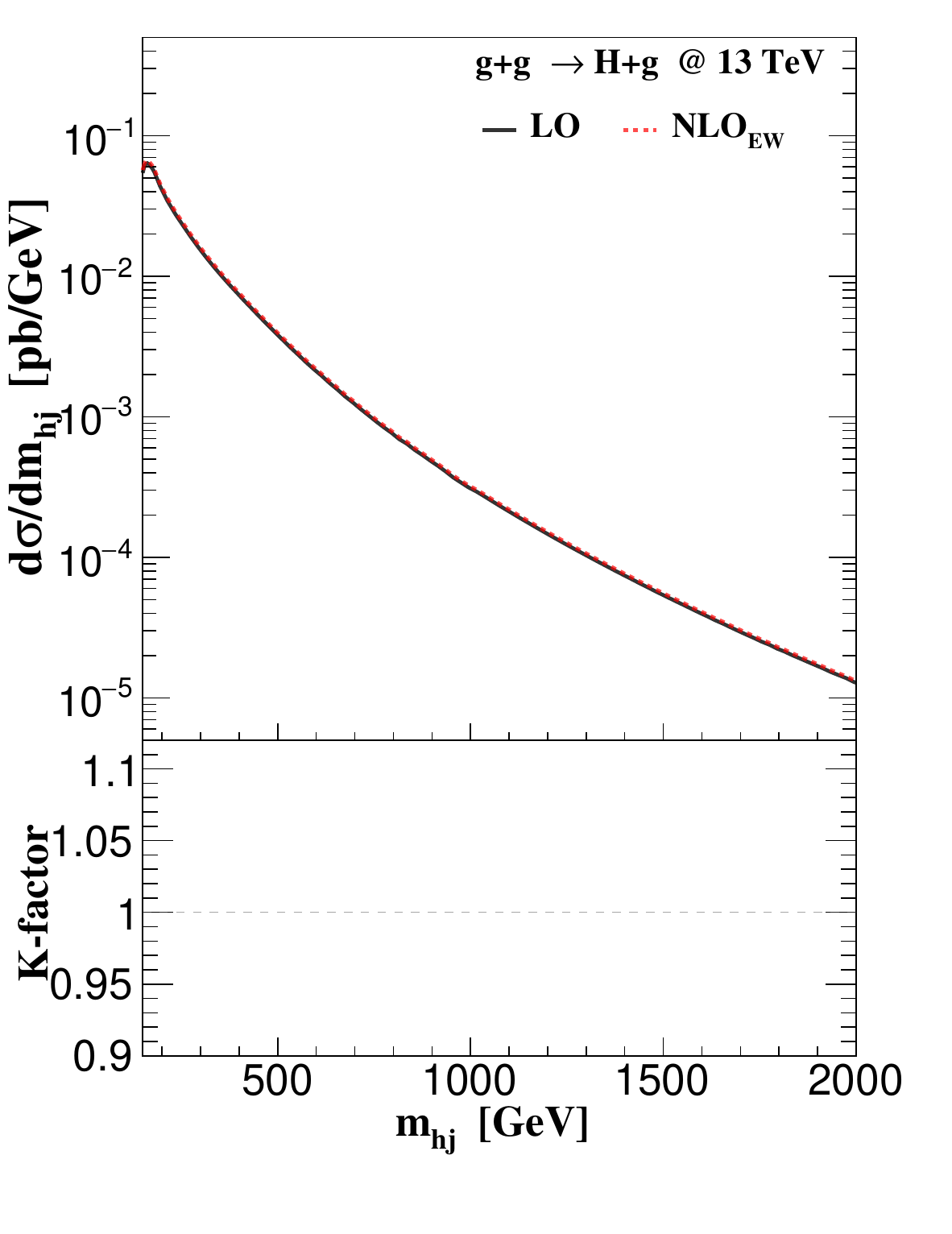}
    \vspace{-0.5cm}
    \caption{Differential cross section with and without NLO EW corrections in $G_{\mu}$ scheme. The upper ones are the differential cross sections and the lower ones are the relative EW corrections as a function of Higgs pseudorapidity $\eta$, Higgs transverse momentum $p_T$, and the invariant mass of Higgs and jet.  }
    \label{fig:diff}
\end{figure}

Fig.~\ref{fig:diff} shows the distributions of Higgs pseudorapidity $\eta$, Higgs transverse momentum $p_T$, and the invariant mass of Higgs and jet.  
The corrections are nearly uniform across the full rapidity range, with the $K$-factor remaining close to 1.04.  The cross section decreases dramatically with increasing $p_T$ or $m_{hj}$. 
The $K$-factor decreases gradually from about 1.05 at low 
$p_T$ to around 0.95 at higher  $p_T$ values, suggesting a negative EW correction in the high-energy regime.
A similar trend is observed as in the $p_T$  case, with the $K$-factor decreasing  mildly as  $m_{hj}$ increases. 
Overall, the EW corrections amount to a few percent effect, and while small, they exhibit non-trivial kinematic dependence—particularly in the high-energy tails of the distributions.

\section{Conclusion}~\label{sec:concl}

In this work, we present the first complete calculation of NLO EW corrections to Higgs boson production in association with a jet via the dominant gluon-gluon fusion channel at the LHC, including the full dependence on the top quark mass. To handle the large number of multi-scale two-loop integrals arising in the calculation, we developed an efficient numerical strategy based on optimized MIs and differential equation methods. The calculation was performed in the on-shell renormalization scheme with three different parameter schemes: $\alpha(0)$, $\alpha(m_Z)$, and $G_\mu$.

Our results show that the EW corrections amount to a few percent enhancement of the total cross section, with a value of approximately 4.3\% in the $G_\mu$ scheme. The differential distributions reveal non-trivial kinematic dependence. Given that the LO amplitudes are explicitly proportional to $G_\mu$, this scheme also ensures a coherent treatment of EW parameters for perturbative studies. 

Our results provide an important step toward improving the precision of Higgs plus jet predictions and reducing theoretical uncertainties in future LHC analyses.

\section* { Acknowledgments}
  
This work was supported in part by the National Natural Science Foundation of China under grant  Nos.   12175048, 12275156, 12321005, and 12375079.
The work of LBC is in part  supported by the Guangdong Basic and Applied Basic Research Foundation  under grant No.  2025B1515020009.
WLS is in part supported by the Natural Science Foundation of ChongQing under Grant No. CSTB2023
NSCQ-MSX0132. 
LBC would also like to thank Yao Ji for computing resource support. 
The authors gratefully acknowledge the valuable discussions and insights provided by the members of the China Collaboration of Precision Testing and New Physics.

{\it Note added:} While this work was under review, an independent study~\cite{Bi:2025oga} reported the same calculation. We have compared the results and found full agreement with our findings.

\bibliography{reference}
\bibliographystyle{JHEP}

\end{document}